\documentclass[aip,pop,reprint]{revtex4-1}
\usepackage[english]{babel}
\usepackage{graphicx}
\usepackage{dcolumn}
\usepackage{bm,color}

\begin{document}

\title{Identifying low-dimensional dynamics in Type-I edge-localised-mode processes in JET plasmas}
\author{F. A. Calderon}
\email{F.A.Calderon-Maldonado@warwick.ac.uk}
\affiliation{Centre for Fusion, Space and Astrophysics, Department of Physics, University of Warwick, Coventry, CV4 7AL, UK}
\author{R. O. Dendy}
\affiliation{EURATOM/CCFE Fusion Association, Culham Science Centre, Abingdon, OX14 3DB, UK}
\affiliation{Centre for Fusion, Space and Astrophysics, Department of Physics, University of Warwick, Coventry, CV4 7AL, UK}
\author{S. C. Chapman}
\affiliation{Centre for Fusion, Space and Astrophysics, Department of Physics, University of Warwick, Coventry, CV4 7AL, UK}
\author{A. J. Webster}
\affiliation{EURATOM/CCFE Fusion Association, Culham Science Centre, Abingdon, OX14 3DB, UK}
\author{B. Alper}
\affiliation{EURATOM/CCFE Fusion Association, Culham Science Centre, Abingdon, OX14 3DB, UK}
\author{R. M. Nicol}
\affiliation{Centre for Fusion, Space and Astrophysics, Department of Physics, University of Warwick, Coventry, CV4 7AL, UK}
\author{JET EFDA Contributors}
\thanks{See the Appendix of F. Romanelli \emph{et al.}, Proceedings of the 23rd IAEA Fusion Energy Conference 2010, Daejeon, Korea.}
\affiliation{JET-EFDA, Culham Science Centre, Abingdon, OX14 3DB, UK}
\date{\today}
\begin{abstract}
Edge localised mode (ELM) measurements from reproducibly similar
plasmas in the Joint European Torus (JET) tokamak, which differ only in their gas puffing rate, are
analysed in terms of the pattern in the sequence of inter-ELM time intervals.
 It is found that the category of ELM defined empirically as Type I - typically more regular, less frequent,
and having larger amplitude than other ELM types - embraces substantially different ELMing
processes.
 By quantifying the structure in the sequence of inter-ELM time intervals using delay time plots, 
we reveal transitions between distinct phase space dynamics, implying transitions between 
distinct underlying physical processes.
 The control parameter for these transitions between these different 
ELMing processes is the gas puffing rate.
\end{abstract}
\maketitle

\section{INTRODUCTION}
Enhanced confinement operating regimes (H-mode) in magnetically
confined plasmas are accompanied by
pulses of energy and particle release known as edge localised modes
(ELMs) \cite{keilhacker84,erckmann93,zohm96,loarte03,kamiya07,Haw2009}.
At steady state, a magnetically confined tokamak plasma comprises a
family of nested magnetic
flux surfaces in a smooth, or laminar state. ELMing constitutes a
relaxation process, for the edge region
of H-mode plasmas, which encompasses an initial trigger for linear MHD
instability evolving into a
fully nonlinear detached state, such that structures propagate to the
first wall where they generate
recombination ra\-dia\-tion. In parallel, local temperature and pressure
gradients evolve rapidly.
The onset of ELMing accompanies a sharp transition in the global state
of the tokamak plasma,
and changes in observed ELM character reflect changes in externally
applied drive such as gas puffing and heating.
Control, mitigation and prediction of the occurrence of large Type I
ELMs are central cha\-llen\-ges for
magnetic confinement fusion plasma physics. There are many active
experimental campaigns in this
area \cite{evans06,liang07,Kirk12}, particularly in support of the future
ITER tokamak, for which the
consequences of uncontrolled Type I ELMs may be unacceptable
\cite{loarte03,Haw2009}. While successful
theo\-ries for some component elements of the ELMing process have been
constructed,
there is currently no comprehensive first principles model that
incorporates all of the physical effects
that are known to contribute to the ELMing process. 
ELM categorisation is primarily
phenomenological \cite{zohm96,loarte03,kamiya07}, furthermore it is
not always easy to discriminate
in real time between Type I and, say, Type III ELMing. Hitherto only a
few papers \cite{degeling01,greenhough03}
have a\-ddres\-sed measured ELM sequences as the pulsed outputs of a
nonlinear system, a
field where generic analysis techniques are well developed and
potential links to ELMing have long
been apparent \cite{itoh91}. Characterisation of ELMing processes by
applying dynamical systems
theo\-ry to the data offers a fresh avenue to understanding, prediction
and control, and may help identify
some of the key properties that models for Type I ELMing must
embody. Here we take the first steps.

D. Ruelle and F. Takens\cite{ruelle71} initiated a classical scenario for the transition from ordered to disordered flow in fluids with increasing driving control parameter\cite{newhouse78,trefethen93}.
This has been observed in Rayleigh-B\'enard convection in fluids \cite{bodenschatz00,alhers74,gollub75,gollub80,libchaber83}, and in drift wave turbulence \cite{klinger97} and flute instabilities in plasmas \cite{brochard06}.
Oscillatory behaviour arises either if there is a constant of the motion, or if there is a limit cycle onto which the system dynamics is a\-ttrac\-ted in the presence of damping or dissipation.
In the present case, where the system is the plasma undergoing the ELMing process, the nature and number of the re\-le\-vant phase space co-ordinates is not known from first principles.
Progress towards their identification can ne\-ver\-the\-less be made by applying techniques of dynamical systems analysis to visualize changes in the topology of the phase space.
A convenient method is that of `delay plots', that is, to plot the successive time intervals between crossings of a surface of section in the phase space\cite{matsoukis00,devine96,schreiber00}.

In this article we report the application of delay plots to the measured time intervals or waiting times between successive ELMs.
We consider ELM sequences from six similar plasmas in the JET tokamak, including JET plasma $57865$ where the H-mode closely approaches an ITER operating regime with respect to some, but not all, key dimensionless parameters \cite{pamela05}.
We obtain evidence that Type I ELMing in these plasmas exhibits transitions between processes with distinct physical analogues, dependent on the value of the gas puffing rate as control pa\-ra\-me\-ter.
In all six plasmas the toroidal magnetic field density is 2.7T, the plasma current is 2.5MA, neutral beam and ion cyclotron resonance heating power are 13.5MW and 2.0MW respectively, and the $H_{98}$ confinement factor is in the range $0.87$ to $1.0$.
 In all six plasmas, gas puffing terminates at $23.3$s and neutral beam heating is ramped down from $23.5$ to $24.5$s.
The differences in Type I ELM character are largely determined by the different levels of externally applied gas puffing.
The intensity of the $D_{\alpha}$ signal, which sometimes saturates, is not necessarily a reliable proxy for the magnitude of the underlying ELM plasma phenomenon, whereas occurrence times are well defined. ELM occurrence and ELM waiting times are the primary physical indicators addressed in the present study. 
The moment of occurrence of each ELM is inferred from the $D_{\alpha}$ datasets using an algorithm similar to that described in \cite{greenhough03},
which exploits the steep leading edge of each ELM. 
This procedure generates a sequence of event times $t_n$ for each $n$th ELM, and hence inter-event times $\delta t_{n}=t_{n}-t_{n-1}$. 
These sequences are used to construct delay plots, which are known\cite{schreiber00,matsoukis00,devine96} to capture aspects of the topology of the unknown underlying phase space evolution of the system.

\begin{figure*}[!htbp]
\begin{center}
\includegraphics[scale=0.245]{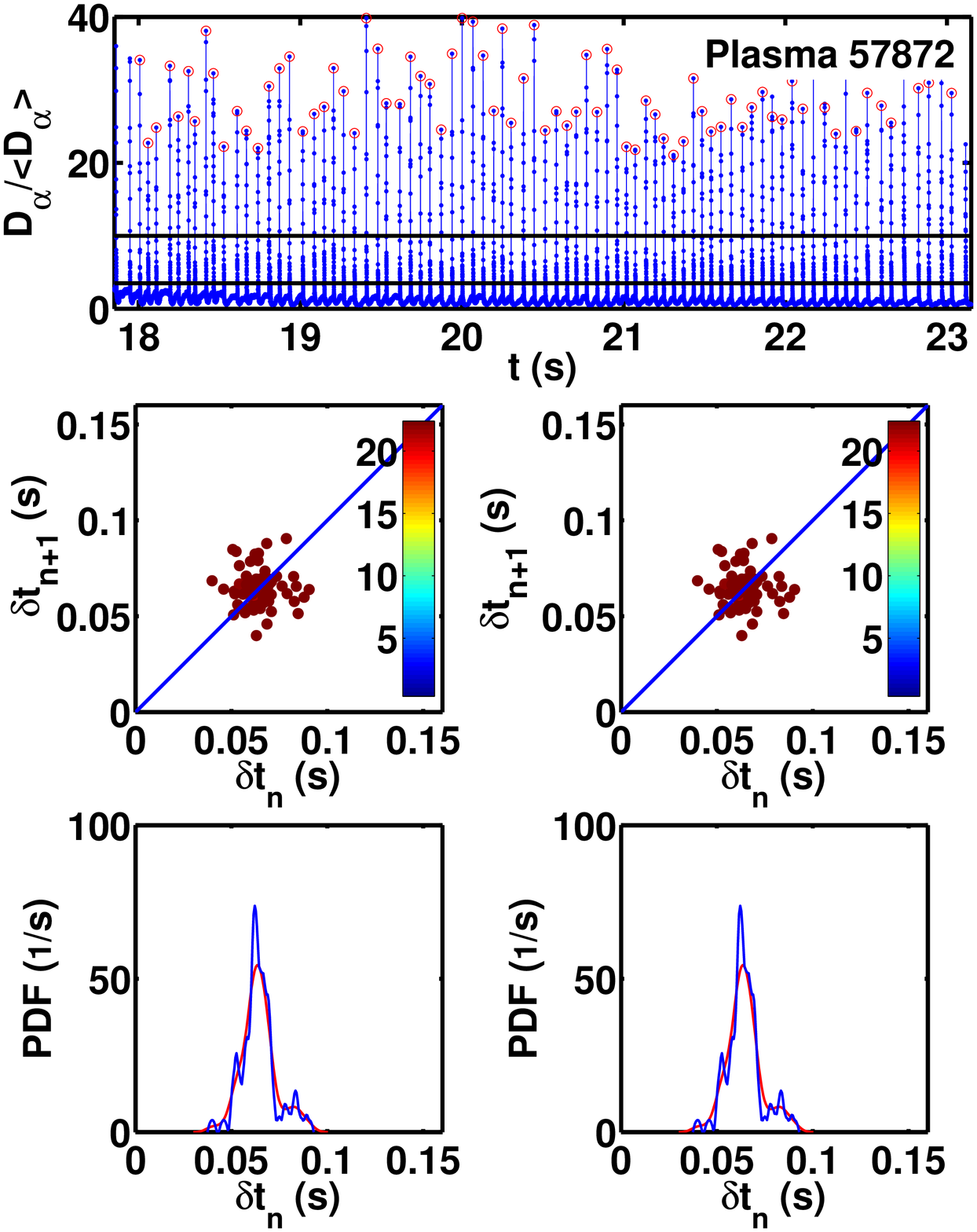}
\includegraphics[scale=0.245]{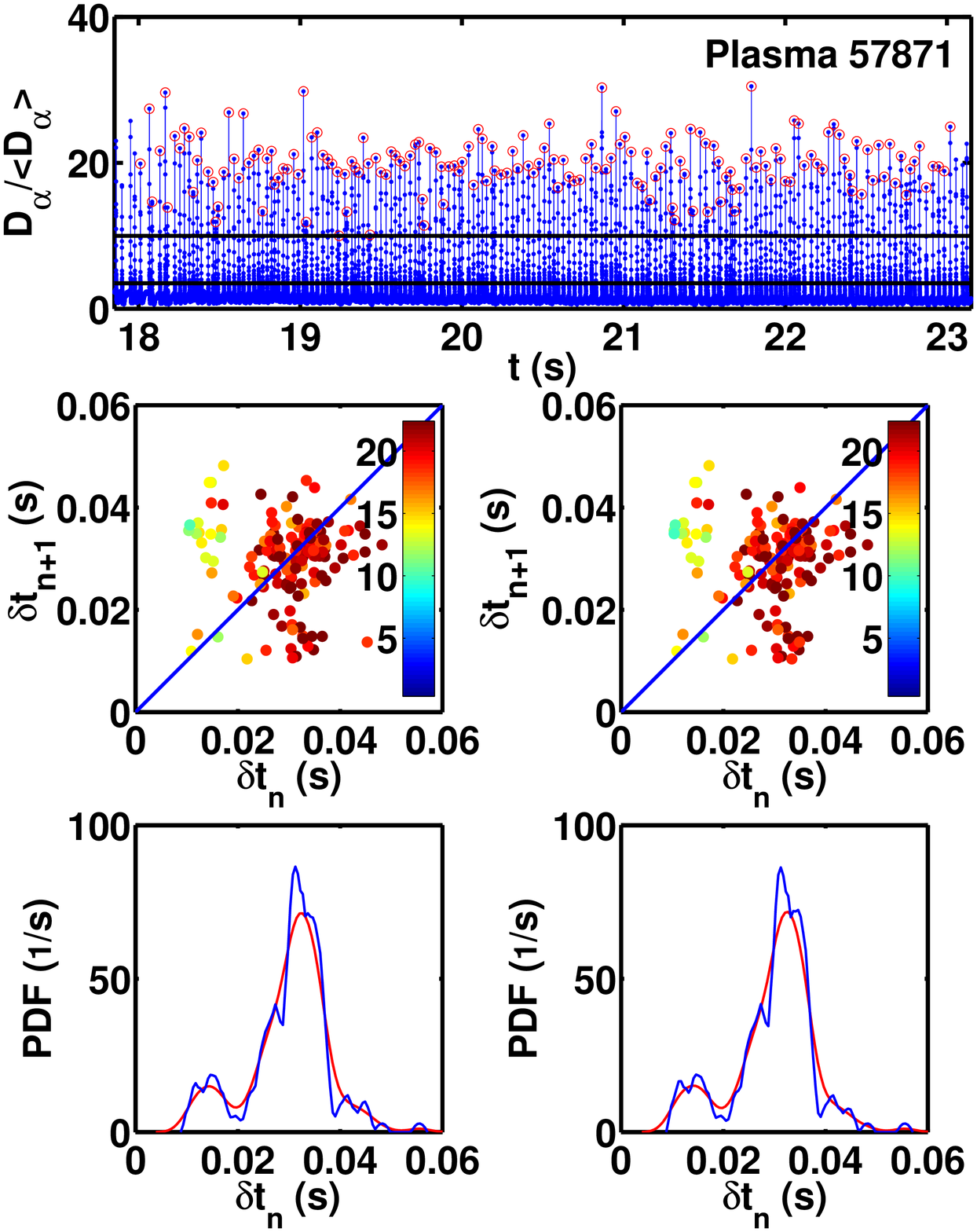}
\includegraphics[scale=0.245]{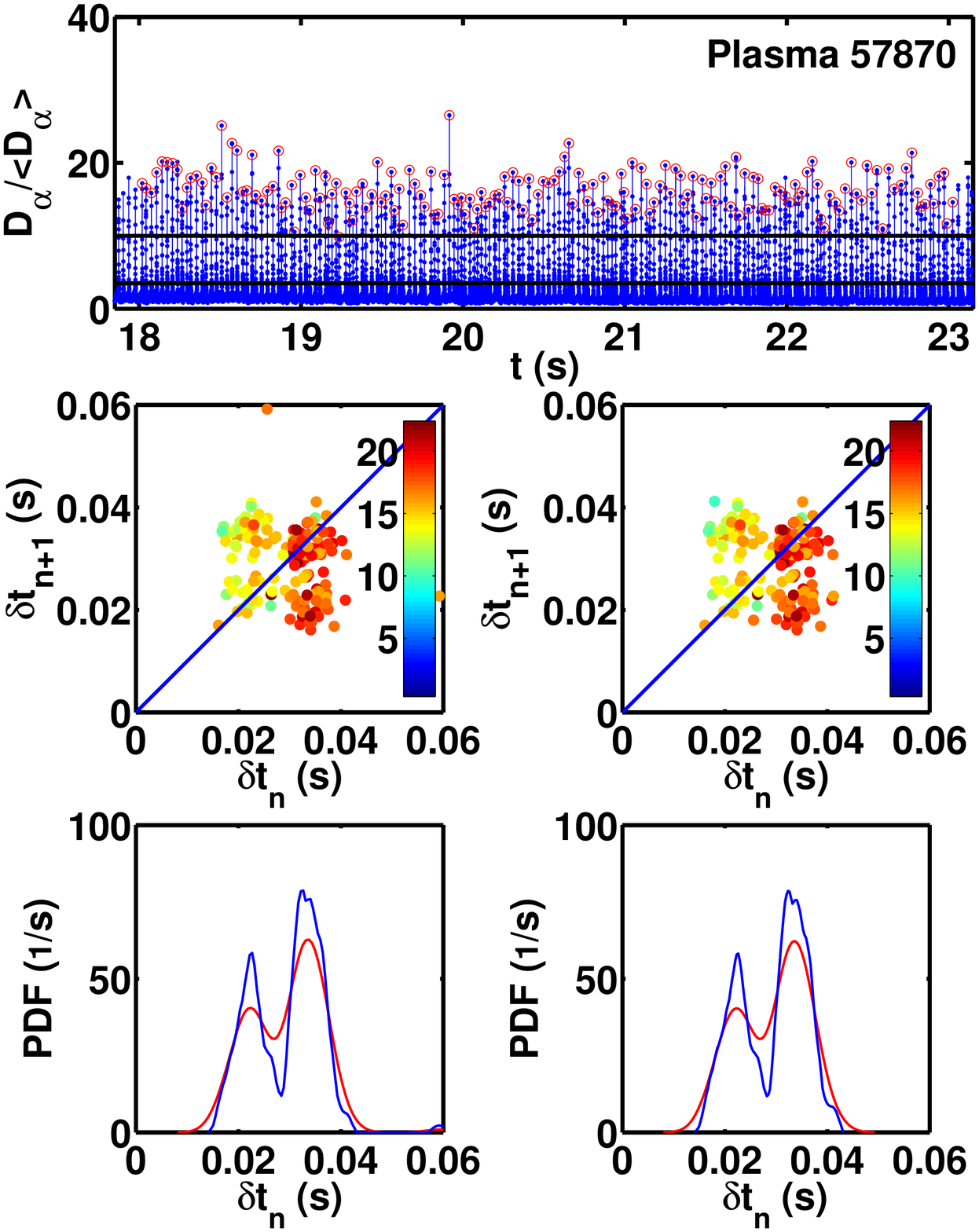}
\caption{\label{Fig.1}
ELM characteristics of three similar JET plasmas $57872$, $57871$,
$57870$ at lower gas puffing rates, showing for each plasma: (top of
each panel) the time trace of $D_{\alpha}$ signal intensity,
displaying also the two amplitude thresholds used for the centre and
bottom plots; (centre of each panel) delay plots for ELMs, with
amplitude colour coded above the higher (lower) threshold on the left
(right); (bottom of each panel) corresponding probability density
functions for the distributions of measured $\delta t_{n}$ for the ELM
time series, using the same amplitude thresholds as for the delay
plots; the red and blue curves represent different binning of the same
data.
The three plasmas are ordered, from the left, in terms of
increasing magnitude of gas puffing, see Fig. 3.}
\end{center}
\end{figure*}
\begin{figure*}[!htbp]
\begin{center}
\includegraphics[scale=0.245]{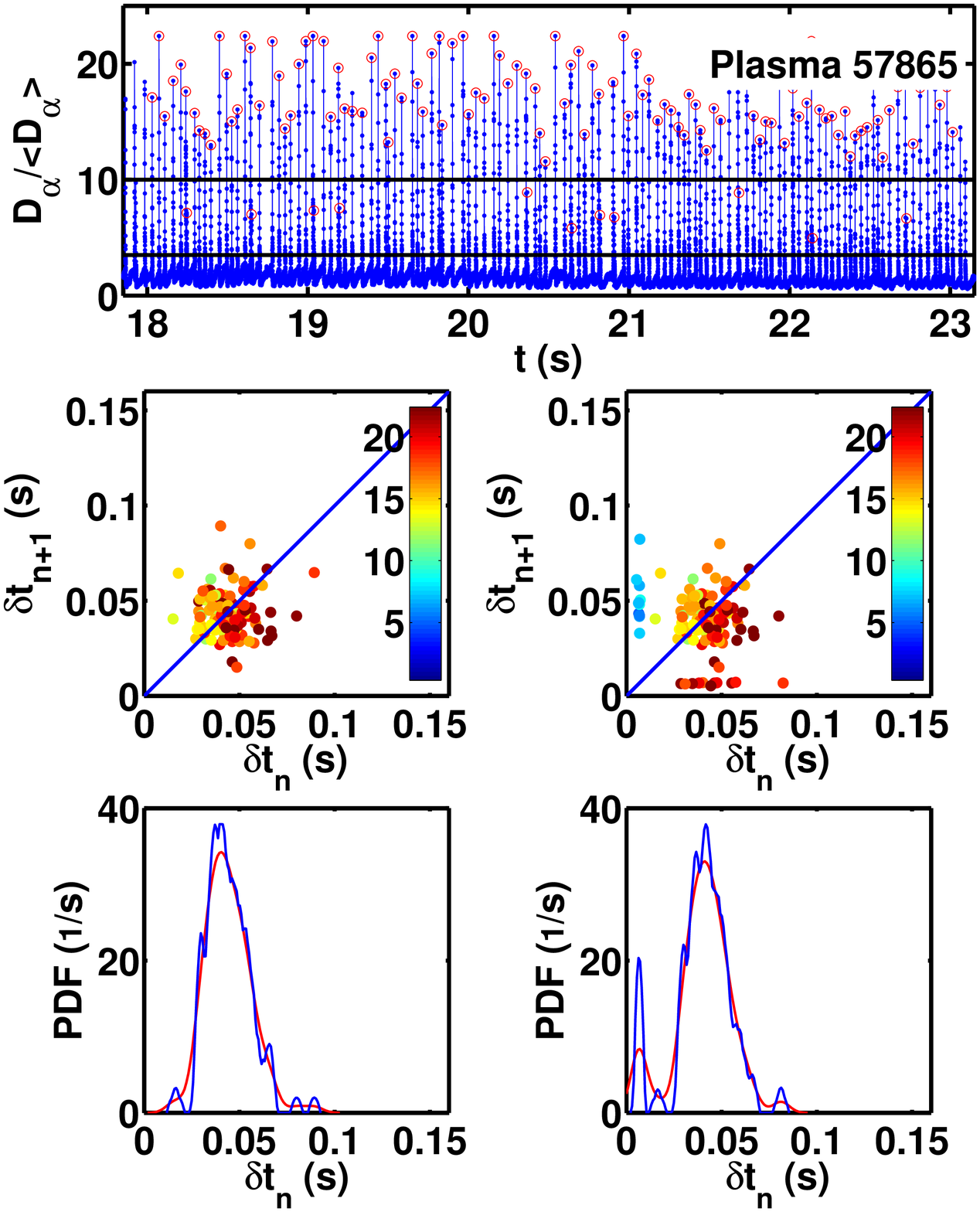}
\includegraphics[scale=0.245]{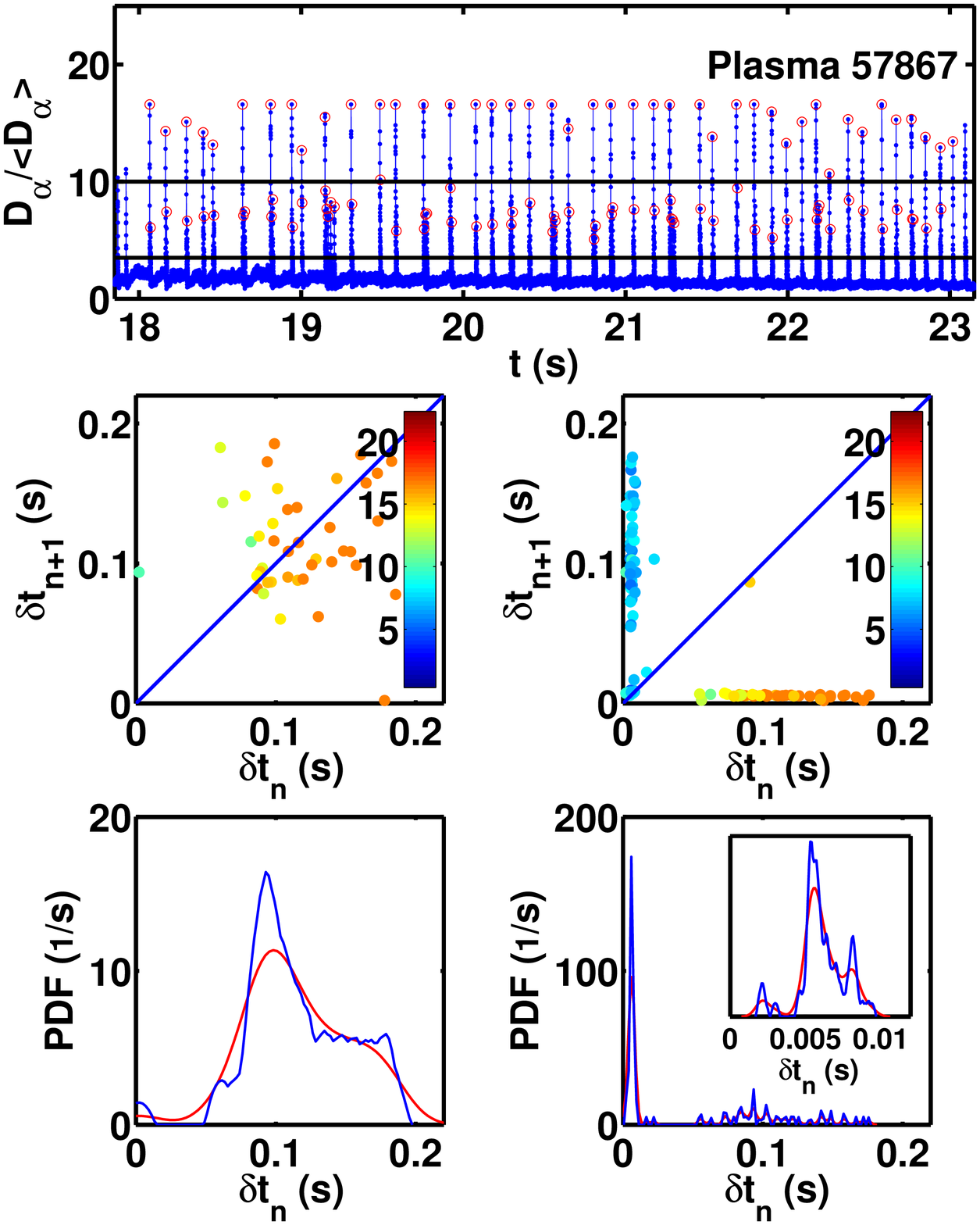}
\includegraphics[scale=0.245]{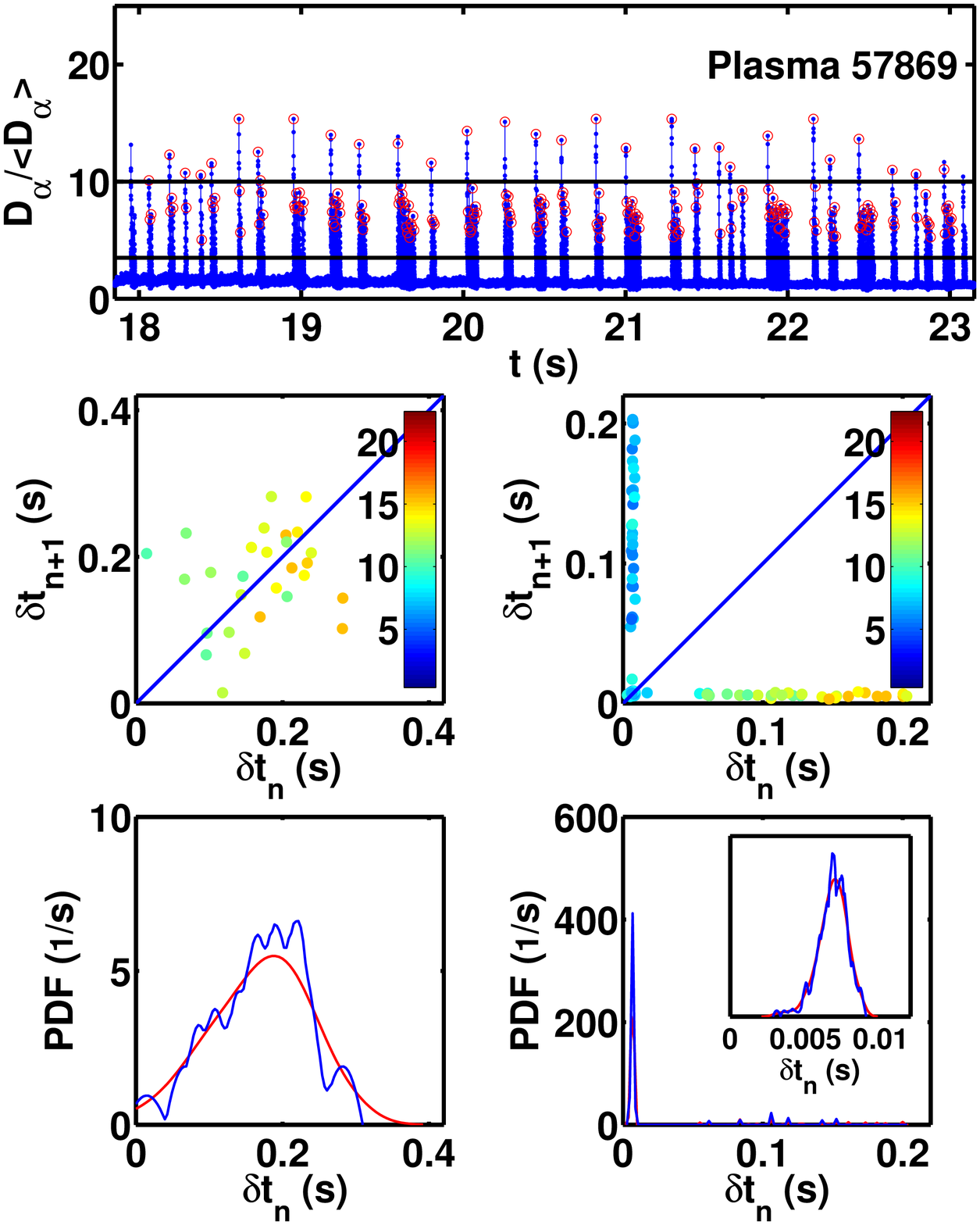}
\caption{\label{Fig.2}
As Fig. 1, for three similar JET plasmas $57865$, $57867$, $57869$ at
higher gas puffing rates. The three plasmas are ordered, from the
left, in terms of increasing magnitude of gas puffing, see Fig. 3. 
The bottom panels from JET plasmas  57867 and 57869 also include an inset panel displaying the sharp peak in the PDF. The population in this sharp peak  increases  with the gas puffing rate, and the average period $\tau= 6.7 \pm 6.6 \times 10^{-2}$ (ms).
}
\end{center}
\end{figure*}
\begin{figure}[!htbp]
\begin{center}
\includegraphics[scale=0.37]{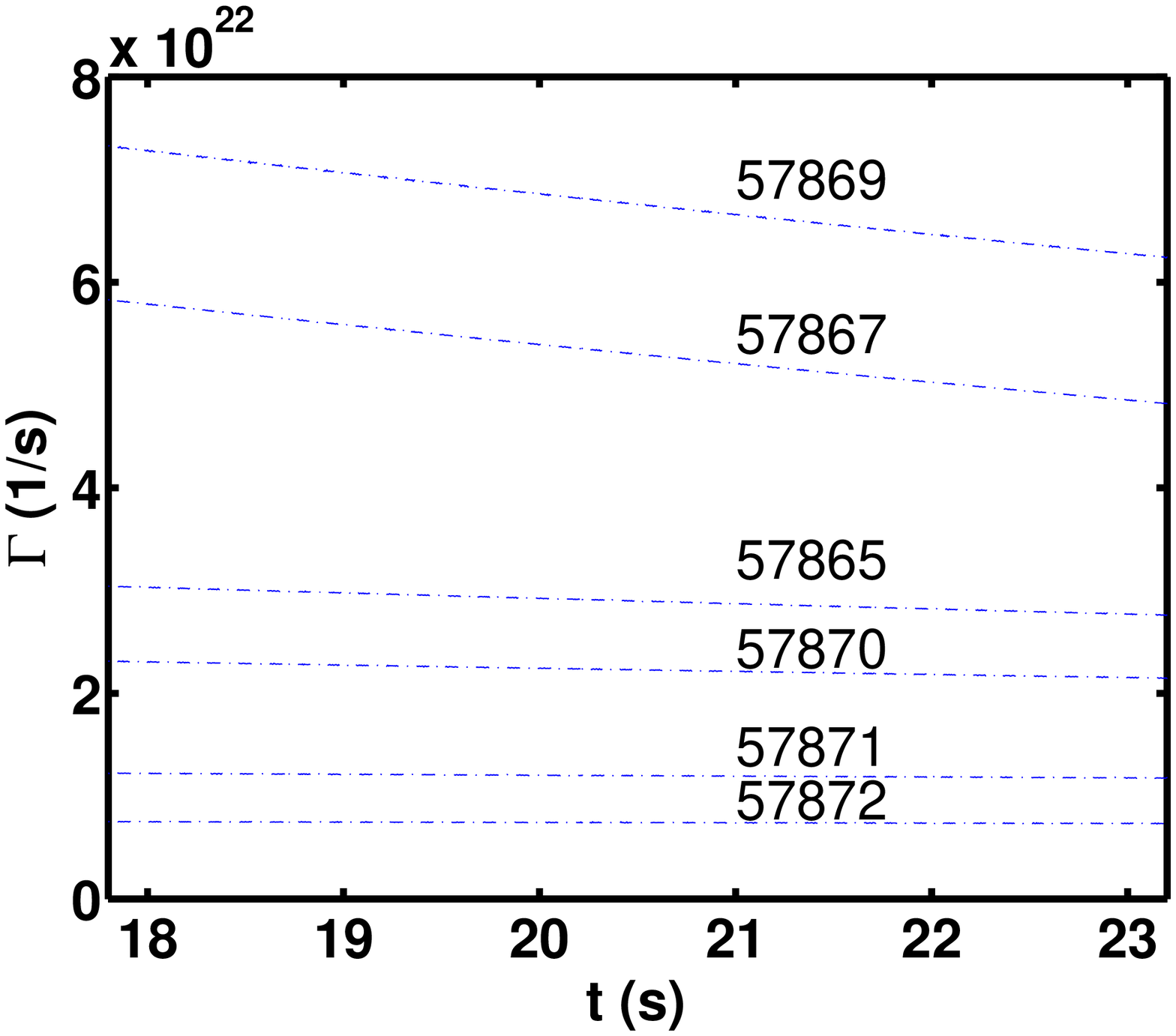}
\caption{\label{Fig.3}
Time trace of gas puffing rate, $\Gamma$, in particles per second, which is the
primary external control
parameter for the six otherwise similar JET plasmas: ordered, from the
bottom, in terms of increasing magnitude.}
\end{center}
\end{figure}
\section{RESULTS AND DISCUSSIONS}

Figures 1 and 2 show measured Type I ELM signals for a sequence of six
JET H-mode plasmas $578nm$, where $nm$ is $72$, $71$, $70$, $65$,
$67$, and $69$ in order of increasing magnitude and duration of the
gas puffing rate, shown in Fig. 3, which is the key external control parameter.
The upper trace in each panel of
Figs. 1 and 2 plots the time-evolving intensity of Lyman alpha
recombination radiation from deuterium, $D_{\alpha}$, measured by a
camera directed at the inner divertor, normalised by the mean
measured intensity.
 The two groupings of three plasmas are at lower
(Fig. 1) and higher (Fig. 2) gas puffing rates. At lower gas puffing rates (Fig. 1)
the ELM signal intensity is roughly the same across each time series, whereas at higher gas
puffing rates (Fig. 2)  this shows a rich structure.
 We will investigate this structure by sorting the ELM events that are used to
construct the time series of inter-ELM time intervals (delay times), in terms of
whether they exceed a threshold in signal intensity; the thresholds
used are indicated by horizontal lines on the ELM time series (top
panel in Figs 1 and 2).
Each $n$th Type I ELM that has signal intensity exceeding a given
threshold then forms a set of events at time $t_n$ with the delay between events $\delta
t_{n}=t_{n}-t_{n-1}$.
The middle panels of Figs. 1 and 2 show the
delay plots for a given thres\-hold, that is, $\delta t_{n+1}$ versus
$\delta t_{n}$.
The $D_{\alpha}$ signal
intensity for the ELM at $t_n$ is indicated by colour coding. 
These delay plots reflect the topology of the system phase space.
For a trajectory
that is approximately singly periodic the
delay plot will exhibit a single
concentration of points on the $\delta t_{n+1}=\delta t_n$ line,
centred on the mean period $\tau=\delta t_{n+1}=\delta t_n$.
The spread of points about the mean period
reflects a combination, in unknown proportions, of intrinsic and
extrinsic sources of irregularity in a quasi-regular process, and
determines the practical resolution limit of this method.
 A period-two
oscillation will generate two concentrations of points, symmetrically
placed either side of the $\delta t_{n+1}=\delta t_n$ line. Dynamical switching between one
period $\tau_1$ and another at $\tau_2$ will generate
four concentrations of points: at the
two distinct periods $\tau_1$ and $\tau_2$  on the $\delta
t_{n+1}=\delta t_n$ line, and at two locations symmetrically
placed either side of the line, at $(\delta t_{n+1},\delta t_n)$ coordinates $(\tau_1,\tau_2)$ and
$(\tau_2,\tau_1)$.

The number of ELMs evaluated in these six JET plasmas ranges between 79 and 197.
 The mean inter-ELM time interval is in the range 25 to 60ms.
The delay plots in Fig. 1 are insensitive to the threshold,  in marked contrast
to Fig. 2, suggesting that these reflect distinct processes.
In Fig. 1, plasmas with successively greater gas puffing rates are
shown from left to right. We can see that increased gas puffing causes the
ELMing process to bifurcate from singly periodic ($57872$), via
transitional behaviour ($57871$), to a situation where two periods are
present ($57870$) together, with the plasma switching between
them.
This behaviour is approximately analogous to that of small
amplitude oscillations of two weakly coupled pendulums with different
natural frequencies. It is also apparent
that a longer delay time $\delta t_n$ before an ELM correlates
statistically with a larger $D_{\alpha}$ signal intensity.
The bottom pair of plots in each panel of Figs. 1 and 2 displays the probability
density functions (pdfs) for the distributions of measured $\delta
t_{n}$ for the ELM time series u\-sing the same amplitude thresholds as
for the delay plots; in Fig. 1, unlike Fig. 2, these two panels are
identical.

We now turn to Fig. 2 which corresponds to higher overall levels of
gas puffing rate. It
displays a  transition in the ELMing process as the
gas puffing rate is increased, which is different to that seen in Fig. 1.
Each ELM with large $D_{\alpha}$ signal intensity is
statistically likely to be rapidly followed by a population of
postcursor ELMs with smaller $D_{\alpha}$ signal intensity. The
likelihood of a postcursor ELM, and their number, increases with gas
puffing rate. As a consequence, the delay plots cons\-truc\-ted for
different thresholds
now, unlike Fig. 1, show different structure. At relatively low gas
puffing rate
(left hand plots) most delays fall within a single group on the
$\delta t_{n+1}=\delta t_n$ line.
However when the threshold is reduced,  smaller postcursor events
begin to feature in the time series of delays and result in
populations (lines parallel to the axes) far from the $\delta
t_{n+1}=\delta t_n$ line, and a new, na\-rrow\-ly constrained group on the
$\delta t_{n+1}=\delta t_n$ line at small $(\delta t_{n+1},\delta
t_n)$. As the
gas puffing rate is increased, these small postcursor events come to
dominate numerically. 
It is noteworthy that whereas ELMs with large signal amplitude exhibit a broad inter-ELM time interval distribution, 
the distribution of the postcursors is very sharply defined and is invariant between the three JET
plasmas, see Fig. 2 bottom panels.
 Its inverse defines a potentially important characteristic frequency
of the ELMing process. This process, as seen in the delay plots, is
analogous to
random large amplitude transient impulses driving  a system that has a
narrowband resonant
frequency response.

Figure 3 displays the gas puffing rates for all six JET plasmas. The
clear changes in ELMing displayed in Fig. 1, and for JET plasmas 57867
to 57869, arise under comparatively small changes in gas puffing,
while there is a relative large step (a factor of approximately two)
between 57865 and 57867. Other ELM
interval dynamics are in principle possible for other gas fuelling rates,
especially for fuelling rates between those of 57865 and 57867, for
these otherwise identical plasma operating regimes.

Some previous experiments have observed that Type I ELM frequency 
(mean inter-ELM interval) increases with gas puffing rate\cite{Macdonald08}. Moreover, early theoretical studies\cite{Lonnroth03} 
suggested that it might be possible to explain the ex\-pe\-ri\-men\-ta\-lly observed
 transition from Type I to Type III ELMy H-mode triggered by strong gas puffing, 
as well as the subsequent increase in ELM frequency and deterioration of plasma 
confinement, as a transition from second to first stability (either ideal or resistive modes). 
However, there is still no widely accepted model for the overall ELMing process or processes.

\section{SUMMARY}

 We have exploited the similarity of these six JET plasmas
 which all have exceptionally long duration $\simeq$
5$s$ of the quasi-stationary ELMing process,  and which
 appear to have only one effective control parameter,
the gas puffing rate. 
These particular experiments yield a sufficient number of ELMs and
 inter-ELM  times, to enable us to  apply the delay plot technique to characterize the dynamics.

There exists an increasing number of ELMing regimes. These are typically characterized phenomenologically and in terms of bulk plasma parameters.
We believe that ELM interval analysis of the kind presented here, if applied more widely, will help shed light on such transitions in confinement phenomenology in tokamak plasmas.
In particular, demonstrating and quantifying the effectiveness of ELM control and mitigation techniques will be assisted by characterizing the measured sequences of inter-ELM time intervals in this way. 

\begin{acknowledgments}
This work, supported by the European Communities under the contract of
Association between EURATOM and CCFE, was carried out within the
framework of the European Fusion Development Agreement. The views and
opinions expressed herein do not necessarily reflect those of the
European Commission. This work was also part-funded by the RCUK Energy
Programme under grant EP/I501045. We acknowledge the UK EPSRC and the Chilean
 committee of science and technology, CO\-NI\-CYT, for support.
\end{acknowledgments}


\begin{thebibliography}{28}
\expandafter\ifx\csname natexlab\endcsname\relax\def\natexlab#1{#1}\fi
\expandafter\ifx\csname bibnamefont\endcsname\relax
  \def\bibnamefont#1{#1}\fi
\expandafter\ifx\csname bibfnamefont\endcsname\relax
  \def\bibfnamefont#1{#1}\fi
\expandafter\ifx\csname citenamefont\endcsname\relax
  \def\citenamefont#1{#1}\fi
\expandafter\ifx\csname url\endcsname\relax
  \def\url#1{\texttt{#1}}\fi
\expandafter\ifx\csname urlprefix\endcsname\relax\def\urlprefix{URL }\fi
\providecommand{\bibinfo}[2]{#2}
\providecommand{\eprint}[2][]{\url{#2}}

\bibitem[{\citenamefont{Keilhacker
  \emph{et~al.}}(1984)\citenamefont{Keilhacker, Becker, Bernhardi, Eberhagen,
  ElShaer, FuBmann, Gehre, Gernhardt, Gierke, Glock
  \emph{et~al.}}}]{keilhacker84}
\bibinfo{author}{\bibfnamefont{M.}~\bibnamefont{Keilhacker}},
  \bibinfo{author}{\bibfnamefont{G.}~\bibnamefont{Becker}},
  \bibinfo{author}{\bibfnamefont{K.}~\bibnamefont{Bernhardi}},
  \bibinfo{author}{\bibfnamefont{A.}~\bibnamefont{Eberhagen}},
  \bibinfo{author}{\bibfnamefont{M.}~\bibnamefont{ElShaer}},
  \bibinfo{author}{\bibfnamefont{G.}~\bibnamefont{FuBmann}},
  \bibinfo{author}{\bibfnamefont{O.}~\bibnamefont{Gehre}},
  \bibinfo{author}{\bibfnamefont{J.}~\bibnamefont{Gernhardt}},
  \bibinfo{author}{\bibfnamefont{G.}~\bibnamefont{Gierke}},
  \bibinfo{author}{\bibfnamefont{E.}~\bibnamefont{Glock}}
  \bibnamefont{\emph{et~al.}}, \bibinfo{journal}{Plasma Phys. Controlled
  Fusion} \textbf{\bibinfo{volume}{26}}, \bibinfo{pages}{49}
  (\bibinfo{year}{1984}).

\bibitem[{\citenamefont{Erckmann \emph{et~al.}}(1993)\citenamefont{Erckmann,
  Wagner, Baldzuhn, Brakel, Burhenn, Gasparino, Grigull, Hartfuss, Hofmann,
  Jaenicke \emph{et~al.}}}]{erckmann93}
\bibinfo{author}{\bibfnamefont{V.}~\bibnamefont{Erckmann}},
  \bibinfo{author}{\bibfnamefont{F.}~\bibnamefont{Wagner}},
  \bibinfo{author}{\bibfnamefont{J.}~\bibnamefont{Baldzuhn}},
  \bibinfo{author}{\bibfnamefont{R.}~\bibnamefont{Brakel}},
  \bibinfo{author}{\bibfnamefont{R.}~\bibnamefont{Burhenn}},
  \bibinfo{author}{\bibfnamefont{U.}~\bibnamefont{Gasparino}},
  \bibinfo{author}{\bibfnamefont{P.}~\bibnamefont{Grigull}},
  \bibinfo{author}{\bibfnamefont{H.~J.} \bibnamefont{Hartfuss}},
  \bibinfo{author}{\bibfnamefont{J.~V.} \bibnamefont{Hofmann}},
  \bibinfo{author}{\bibfnamefont{R.}~\bibnamefont{Jaenicke}}
  \bibnamefont{\emph{et~al.}}, \bibinfo{journal}{Phys. Rev. Lett.}
  \textbf{\bibinfo{volume}{70}}, \bibinfo{pages}{2086} (\bibinfo{year}{1993}).

\bibitem[{\citenamefont{Zohm}(1996)}]{zohm96}
\bibinfo{author}{\bibfnamefont{H.}~\bibnamefont{Zohm}},
  \bibinfo{journal}{Plasma Phys. Controlled Fusion}
  \textbf{\bibinfo{volume}{38}}, \bibinfo{pages}{105} (\bibinfo{year}{1996}).

\bibitem[{\citenamefont{Loarte \emph{et~al.}}(2003)\citenamefont{Loarte,
  Saibene, Sartori, Campbell, Becoulet, Horton, Eich, Herrmann, Matthews,
  Asakura \emph{et~al.}}}]{loarte03}
\bibinfo{author}{\bibfnamefont{A.}~\bibnamefont{Loarte}},
  \bibinfo{author}{\bibfnamefont{G.}~\bibnamefont{Saibene}},
  \bibinfo{author}{\bibfnamefont{R.}~\bibnamefont{Sartori}},
  \bibinfo{author}{\bibfnamefont{D.}~\bibnamefont{Campbell}},
  \bibinfo{author}{\bibfnamefont{M.}~\bibnamefont{Becoulet}},
  \bibinfo{author}{\bibfnamefont{L.}~\bibnamefont{Horton}},
  \bibinfo{author}{\bibfnamefont{T.}~\bibnamefont{Eich}},
  \bibinfo{author}{\bibfnamefont{A.}~\bibnamefont{Herrmann}},
  \bibinfo{author}{\bibfnamefont{G.}~\bibnamefont{Matthews}},
  \bibinfo{author}{\bibfnamefont{N.}~\bibnamefont{Asakura}}
  \bibnamefont{\emph{et~al.}}, \bibinfo{journal}{Plasma Phys. Controlled
  Fusion} \textbf{\bibinfo{volume}{45}}, \bibinfo{pages}{1549}
  (\bibinfo{year}{2003}).

\bibitem[{\citenamefont{Kamiya \emph{et~al.}}(2007)\citenamefont{Kamiya,
  Asakura, Boedo, Eich, Federici, Fenstermacher, Finken, Herrmann, Terry, kirk
  \emph{et~al.}}}]{kamiya07}
\bibinfo{author}{\bibfnamefont{K.}~\bibnamefont{Kamiya}},
  \bibinfo{author}{\bibfnamefont{N.}~\bibnamefont{Asakura}},
  \bibinfo{author}{\bibfnamefont{J.}~\bibnamefont{Boedo}},
  \bibinfo{author}{\bibfnamefont{T.}~\bibnamefont{Eich}},
  \bibinfo{author}{\bibfnamefont{G.}~\bibnamefont{Federici}},
  \bibinfo{author}{\bibfnamefont{M.}~\bibnamefont{Fenstermacher}},
  \bibinfo{author}{\bibfnamefont{K.}~\bibnamefont{Finken}},
  \bibinfo{author}{\bibfnamefont{A.}~\bibnamefont{Herrmann}},
  \bibinfo{author}{\bibfnamefont{J.}~\bibnamefont{Terry}},
  \bibinfo{author}{\bibfnamefont{A.}~\bibnamefont{kirk}}
  \bibnamefont{\emph{et~al.}}, \bibinfo{journal}{Plasma Phys. Controlled
  Fusion} \textbf{\bibinfo{volume}{49}}, \bibinfo{pages}{S43}
  (\bibinfo{year}{2007}).

\bibitem[{\citenamefont{Hawryluk \emph{et~al.}}(2009)\citenamefont{Hawryluk,
  Campbell, Janeschitz, Thomas, Albanese, Ambrosino, Bachmann, Baylor,
  Becoulet, Benfatto \emph{et~al.}}}]{Haw2009}
\bibinfo{author}{\bibfnamefont{R.~J.} \bibnamefont{Hawryluk}},
  \bibinfo{author}{\bibfnamefont{D.~J.} \bibnamefont{Campbell}},
  \bibinfo{author}{\bibfnamefont{G.}~\bibnamefont{Janeschitz}},
  \bibinfo{author}{\bibfnamefont{P.~R.} \bibnamefont{Thomas}},
  \bibinfo{author}{\bibfnamefont{R.}~\bibnamefont{Albanese}},
  \bibinfo{author}{\bibfnamefont{R.}~\bibnamefont{Ambrosino}},
  \bibinfo{author}{\bibfnamefont{C.}~\bibnamefont{Bachmann}},
  \bibinfo{author}{\bibfnamefont{L.}~\bibnamefont{Baylor}},
  \bibinfo{author}{\bibfnamefont{M.}~\bibnamefont{Becoulet}},
  \bibinfo{author}{\bibfnamefont{I.}~\bibnamefont{Benfatto}}
  \bibnamefont{\emph{et~al.}}, \bibinfo{journal}{Nucl. Fusion}
  \textbf{\bibinfo{volume}{49}}, \bibinfo{pages}{065012}
  (\bibinfo{year}{2009}).

\bibitem[{\citenamefont{Evans \emph{et~al.}}(2006)\citenamefont{Evans, Moyer,
  Burrell, Fenstermacher, Joseph, Leonard, Osborne, Porter, Schaffer, Snyder
  \emph{et~al.}}}]{evans06}
\bibinfo{author}{\bibfnamefont{T.~E.} \bibnamefont{Evans}},
  \bibinfo{author}{\bibfnamefont{R.~A.} \bibnamefont{Moyer}},
  \bibinfo{author}{\bibfnamefont{K.~H.} \bibnamefont{Burrell}},
  \bibinfo{author}{\bibfnamefont{M.~E.} \bibnamefont{Fenstermacher}},
  \bibinfo{author}{\bibfnamefont{I.}~\bibnamefont{Joseph}},
  \bibinfo{author}{\bibfnamefont{A.~W.} \bibnamefont{Leonard}},
  \bibinfo{author}{\bibfnamefont{T.~H.} \bibnamefont{Osborne}},
  \bibinfo{author}{\bibfnamefont{G.~D.} \bibnamefont{Porter}},
  \bibinfo{author}{\bibfnamefont{M.~J.} \bibnamefont{Schaffer}},
  \bibinfo{author}{\bibfnamefont{P.~B.} \bibnamefont{Snyder}}
  \bibnamefont{\emph{et~al.}}, \bibinfo{journal}{Nat. Phys.}
  \textbf{\bibinfo{volume}{2}}, \bibinfo{pages}{419} (\bibinfo{year}{2006}).

\bibitem[{\citenamefont{Liang \emph{et~al.}}(2007)\citenamefont{Liang,
  Koslowski, Thomas, Nardon, Alper, Andrew, Andrew, Arnoux, Baranov,
  B{\'e}coulet \emph{et~al.}}}]{liang07}
\bibinfo{author}{\bibfnamefont{Y.}~\bibnamefont{Liang}},
  \bibinfo{author}{\bibfnamefont{H.~R.} \bibnamefont{Koslowski}},
  \bibinfo{author}{\bibfnamefont{P.~R.} \bibnamefont{Thomas}},
  \bibinfo{author}{\bibfnamefont{E.}~\bibnamefont{Nardon}},
  \bibinfo{author}{\bibfnamefont{B.}~\bibnamefont{Alper}},
  \bibinfo{author}{\bibfnamefont{P.}~\bibnamefont{Andrew}},
  \bibinfo{author}{\bibfnamefont{Y.}~\bibnamefont{Andrew}},
  \bibinfo{author}{\bibfnamefont{G.}~\bibnamefont{Arnoux}},
  \bibinfo{author}{\bibfnamefont{Y.}~\bibnamefont{Baranov}},
  \bibinfo{author}{\bibfnamefont{M.}~\bibnamefont{B{\'e}coulet}}
  \bibnamefont{\emph{et~al.}}, \bibinfo{journal}{Phys. Rev. Lett.}
  \textbf{\bibinfo{volume}{98}}, \bibinfo{pages}{265004}
  (\bibinfo{year}{2007}).

\bibitem[{\citenamefont{Kirk \emph{et~al.}}(2012)\citenamefont{Kirk, Harrison,
  Liu, Nardon, Chapman, and Denner}}]{Kirk12}
\bibinfo{author}{\bibfnamefont{A.}~\bibnamefont{Kirk}},
  \bibinfo{author}{\bibfnamefont{J.}~\bibnamefont{Harrison}},
  \bibinfo{author}{\bibfnamefont{Y.}~\bibnamefont{Liu}},
  \bibinfo{author}{\bibfnamefont{E.}~\bibnamefont{Nardon}},
  \bibinfo{author}{\bibfnamefont{I.~T.} \bibnamefont{Chapman}}
  \bibnamefont{and} \bibinfo{author}{\bibfnamefont{P.}~\bibnamefont{Denner}}
  (\bibinfo{collaboration}{the MAST team}), \bibinfo{journal}{Phys. Rev. Lett.}
  \textbf{\bibinfo{volume}{108}}, \bibinfo{pages}{255003}
  (\bibinfo{year}{2012}).

\bibitem[{\citenamefont{Degeling \emph{et~al.}}(2001)\citenamefont{Degeling,
  Martin, Bak, Lister, and Llobet}}]{degeling01}
\bibinfo{author}{\bibfnamefont{A.~W.} \bibnamefont{Degeling}},
  \bibinfo{author}{\bibfnamefont{Y.~R.} \bibnamefont{Martin}},
  \bibinfo{author}{\bibfnamefont{P.~E.} \bibnamefont{Bak}},
  \bibinfo{author}{\bibfnamefont{J.~B.} \bibnamefont{Lister}} \bibnamefont{and}
  \bibinfo{author}{\bibfnamefont{X.}~\bibnamefont{Llobet}},
  \bibinfo{journal}{Plasma Phys. Controlled Fusion}
  \textbf{\bibinfo{volume}{43}}, \bibinfo{pages}{1671} (\bibinfo{year}{2001}).

\bibitem[{\citenamefont{Greenhough
  \emph{et~al.}}(2003)\citenamefont{Greenhough, Chapman, Dendy, and
  Ward}}]{greenhough03}
\bibinfo{author}{\bibfnamefont{J.}~\bibnamefont{Greenhough}},
  \bibinfo{author}{\bibfnamefont{S.~C.} \bibnamefont{Chapman}},
  \bibinfo{author}{\bibfnamefont{R.~O.} \bibnamefont{Dendy}} \bibnamefont{and}
  \bibinfo{author}{\bibfnamefont{D.~J.} \bibnamefont{Ward}},
  \bibinfo{journal}{Plasma Phys. Controlled Fusion}
  \textbf{\bibinfo{volume}{45}}, \bibinfo{pages}{747} (\bibinfo{year}{2003}).

\bibitem[{\citenamefont{Itoh \emph{et~al.}}(1991)\citenamefont{Itoh, Itoh,
  Fukuyama, and Miura}}]{itoh91}
\bibinfo{author}{\bibfnamefont{S.-I.} \bibnamefont{Itoh}},
  \bibinfo{author}{\bibfnamefont{K.}~\bibnamefont{Itoh}},
  \bibinfo{author}{\bibfnamefont{A.}~\bibnamefont{Fukuyama}} \bibnamefont{and}
  \bibinfo{author}{\bibfnamefont{Y.}~\bibnamefont{Miura}},
  \bibinfo{journal}{Phys. Rev. Lett.} \textbf{\bibinfo{volume}{67}},
  \bibinfo{pages}{2485} (\bibinfo{year}{1991}).

\bibitem[{\citenamefont{Ruelle and Takens}(1971)}]{ruelle71}
\bibinfo{author}{\bibfnamefont{D.}~\bibnamefont{Ruelle}} \bibnamefont{and}
  \bibinfo{author}{\bibfnamefont{F.}~\bibnamefont{Takens}},
  \bibinfo{journal}{Commun. Math. Phys.} \textbf{\bibinfo{volume}{20}},
  \bibinfo{pages}{167} (\bibinfo{year}{1971}).

\bibitem[{\citenamefont{Newhouse \emph{et~al.}}(1978)\citenamefont{Newhouse,
  Ruelle, and Takens}}]{newhouse78}
\bibinfo{author}{\bibfnamefont{S.}~\bibnamefont{Newhouse}},
  \bibinfo{author}{\bibfnamefont{D.}~\bibnamefont{Ruelle}} \bibnamefont{and}
  \bibinfo{author}{\bibfnamefont{F.}~\bibnamefont{Takens}},
  \bibinfo{journal}{Commun. Math. Phys.} \textbf{\bibinfo{volume}{64}},
  \bibinfo{pages}{35} (\bibinfo{year}{1978}).

\bibitem[{\citenamefont{Trefethen \emph{et~al.}}(1993)\citenamefont{Trefethen,
  Trefethen, Reddy, and Driscoll}}]{trefethen93}
\bibinfo{author}{\bibfnamefont{L.~N.} \bibnamefont{Trefethen}},
  \bibinfo{author}{\bibfnamefont{A.~E.} \bibnamefont{Trefethen}},
  \bibinfo{author}{\bibfnamefont{S.~C.} \bibnamefont{Reddy}} \bibnamefont{and}
  \bibinfo{author}{\bibfnamefont{T.~A.} \bibnamefont{Driscoll}},
  \bibinfo{journal}{Science} \textbf{\bibinfo{volume}{261}},
  \bibinfo{pages}{578} (\bibinfo{year}{1993}).

\bibitem[{\citenamefont{Bodenschatz
  \emph{et~al.}}(2000)\citenamefont{Bodenschatz, Pesch, and
  Ahlers}}]{bodenschatz00}
\bibinfo{author}{\bibfnamefont{E.}~\bibnamefont{Bodenschatz}},
  \bibinfo{author}{\bibfnamefont{W.}~\bibnamefont{Pesch}} \bibnamefont{and}
  \bibinfo{author}{\bibfnamefont{G.}~\bibnamefont{Ahlers}},
  \bibinfo{journal}{Annu. Rev. Fluid Mech.} \textbf{\bibinfo{volume}{32}},
  \bibinfo{pages}{709} (\bibinfo{year}{2000}).

\bibitem[{\citenamefont{Ahlers}(1974)}]{alhers74}
\bibinfo{author}{\bibfnamefont{G.}~\bibnamefont{Ahlers}},
  \bibinfo{journal}{Phys. Rev. Lett.} \textbf{\bibinfo{volume}{33}},
  \bibinfo{pages}{1185} (\bibinfo{year}{1974}).

\bibitem[{\citenamefont{Gollub and Swinney}(1975)}]{gollub75}
\bibinfo{author}{\bibfnamefont{J.~P.} \bibnamefont{Gollub}} \bibnamefont{and}
  \bibinfo{author}{\bibfnamefont{H.~L.} \bibnamefont{Swinney}},
  \bibinfo{journal}{Phys. Rev. Lett.} \textbf{\bibinfo{volume}{35}},
  \bibinfo{pages}{927} (\bibinfo{year}{1975}).

\bibitem[{\citenamefont{Gollub and Benson}(1980)}]{gollub80}
\bibinfo{author}{\bibfnamefont{J.~P.} \bibnamefont{Gollub}} \bibnamefont{and}
  \bibinfo{author}{\bibfnamefont{S.~V.} \bibnamefont{Benson}},
  \bibinfo{journal}{J. Fluid Mech.} \textbf{\bibinfo{volume}{100}},
  \bibinfo{pages}{449} (\bibinfo{year}{1980}).

\bibitem[{\citenamefont{Libchaber \emph{et~al.}}(1983)\citenamefont{Libchaber,
  Fauve, and Laroche}}]{libchaber83}
\bibinfo{author}{\bibfnamefont{A.}~\bibnamefont{Libchaber}},
  \bibinfo{author}{\bibfnamefont{S.}~\bibnamefont{Fauve}} \bibnamefont{and}
  \bibinfo{author}{\bibfnamefont{C.}~\bibnamefont{Laroche}},
  \bibinfo{journal}{Physica D} \textbf{\bibinfo{volume}{7}},
  \bibinfo{pages}{73} (\bibinfo{year}{1983}).

\bibitem[{\citenamefont{Klinger \emph{et~al.}}(1997)\citenamefont{Klinger,
  Latten, Piel, Bonhomme, Pierre, and {Dudok de Wit}}}]{klinger97}
\bibinfo{author}{\bibfnamefont{T.}~\bibnamefont{Klinger}},
  \bibinfo{author}{\bibfnamefont{A.}~\bibnamefont{Latten}},
  \bibinfo{author}{\bibfnamefont{A.}~\bibnamefont{Piel}},
  \bibinfo{author}{\bibfnamefont{G.}~\bibnamefont{Bonhomme}},
  \bibinfo{author}{\bibfnamefont{T.}~\bibnamefont{Pierre}} \bibnamefont{and}
  \bibinfo{author}{\bibfnamefont{T.}~\bibnamefont{{Dudok de Wit}}},
  \bibinfo{journal}{Phys. Rev. Lett.} \textbf{\bibinfo{volume}{79}},
  \bibinfo{pages}{3913} (\bibinfo{year}{1997}).

\bibitem[{\citenamefont{Brochard \emph{et~al.}}(2006)\citenamefont{Brochard,
  Gravier, and Bonhomme}}]{brochard06}
\bibinfo{author}{\bibfnamefont{F.}~\bibnamefont{Brochard}},
  \bibinfo{author}{\bibfnamefont{E.}~\bibnamefont{Gravier}} \bibnamefont{and}
  \bibinfo{author}{\bibfnamefont{G.}~\bibnamefont{Bonhomme}},
  \bibinfo{journal}{Phys. Rev. E} \textbf{\bibinfo{volume}{73}},
  \bibinfo{pages}{036403} (\bibinfo{year}{2006}).

\bibitem[{\citenamefont{Matsoukis \emph{et~al.}}(2000)\citenamefont{Matsoukis,
  Chapman, and Rowlands}}]{matsoukis00}
\bibinfo{author}{\bibfnamefont{S.~K.} \bibnamefont{Matsoukis}},
  \bibinfo{author}{\bibfnamefont{S.}~\bibnamefont{Chapman}} \bibnamefont{and}
  \bibinfo{author}{\bibfnamefont{G.}~\bibnamefont{Rowlands}},
  \bibinfo{journal}{Physica D} \textbf{\bibinfo{volume}{138}},
  \bibinfo{pages}{251 } (\bibinfo{year}{2000}).

\bibitem[{\citenamefont{Devine and Chapman}(1996)}]{devine96}
\bibinfo{author}{\bibfnamefont{P.~E.} \bibnamefont{Devine}} \bibnamefont{and}
  \bibinfo{author}{\bibfnamefont{S.~C.} \bibnamefont{Chapman}},
  \bibinfo{journal}{Physica D} \textbf{\bibinfo{volume}{95}},
  \bibinfo{pages}{35 } (\bibinfo{year}{1996}).

\bibitem[{\citenamefont{Schreiber and Schmitz}(2000)}]{schreiber00}
\bibinfo{author}{\bibfnamefont{T.}~\bibnamefont{Schreiber}} \bibnamefont{and}
  \bibinfo{author}{\bibfnamefont{A.}~\bibnamefont{Schmitz}},
  \bibinfo{journal}{Physica D}
  \textbf{\bibinfo{volume}{142}}, \bibinfo{pages}{346 } (\bibinfo{year}{2000}).

\bibitem[{\citenamefont{Pamela \emph{et~al.}}(2005)\citenamefont{Pamela,
  Ongena, and Contributors}}]{pamela05}
\bibinfo{author}{\bibfnamefont{J.}~\bibnamefont{Pamela}},
  \bibinfo{author}{\bibfnamefont{J.}~\bibnamefont{Ongena}} \bibnamefont{and}
  \bibinfo{author}{\bibfnamefont{J.~E.} \bibnamefont{Contributors}},
  \bibinfo{journal}{Nucl. Fusion} \textbf{\bibinfo{volume}{45}},
  \bibinfo{pages}{S63} (\bibinfo{year}{2005}).

\bibitem[{\citenamefont{McDonald \emph{et~al.}}(2008)\citenamefont{McDonald,
  Andrew, Huysmans, Loarte, Ongena, Rapp, and Saarelma}}]{Macdonald08}
\bibinfo{author}{\bibfnamefont{D.~C.} \bibnamefont{McDonald}},
  \bibinfo{author}{\bibfnamefont{Y.}~\bibnamefont{Andrew}},
  \bibinfo{author}{\bibfnamefont{G.~T.~A.} \bibnamefont{Huysmans}},
  \bibinfo{author}{\bibfnamefont{A.}~\bibnamefont{Loarte}},
  \bibinfo{author}{\bibfnamefont{J.}~\bibnamefont{Ongena}},
  \bibinfo{author}{\bibfnamefont{J.}~\bibnamefont{Rapp}} \bibnamefont{and}
  \bibinfo{author}{\bibfnamefont{S.}~\bibnamefont{Saarelma}},
  \bibinfo{journal}{Fusion Sci. Technol.} \textbf{\bibinfo{volume}{53}},
  \bibinfo{pages}{891} (\bibinfo{year}{2008}).

\bibitem[{\citenamefont{L$\mbox{\"{o}}$nnroth
  \emph{et~al.}}(2003)\citenamefont{L$\mbox{\"{o}}$nnroth, Parail, Corrigan,
  Heading, Huysmans, Loarte, Saarelma, Saibene, Sharapov, Spence
  \emph{et~al.}}}]{Lonnroth03}
\bibinfo{author}{\bibfnamefont{J.-S.} \bibnamefont{L$\mbox{\"{o}}$nnroth}},
  \bibinfo{author}{\bibfnamefont{V.~V.} \bibnamefont{Parail}},
  \bibinfo{author}{\bibfnamefont{G.}~\bibnamefont{Corrigan}},
  \bibinfo{author}{\bibfnamefont{D.}~\bibnamefont{Heading}},
  \bibinfo{author}{\bibfnamefont{G.}~\bibnamefont{Huysmans}},
  \bibinfo{author}{\bibfnamefont{A.}~\bibnamefont{Loarte}},
  \bibinfo{author}{\bibfnamefont{S.}~\bibnamefont{Saarelma}},
  \bibinfo{author}{\bibfnamefont{G.}~\bibnamefont{Saibene}},
  \bibinfo{author}{\bibfnamefont{S.}~\bibnamefont{Sharapov}},
  \bibinfo{author}{\bibfnamefont{J.}~\bibnamefont{Spence}}
  \bibnamefont{\emph{et~al.}}, \bibinfo{journal}{Plasma Phys. Controlled
  Fusion} \textbf{\bibinfo{volume}{45}}, \bibinfo{pages}{1689}
  (\bibinfo{year}{2003}).

\end{thebibliography}
\end{document}